\documentclass[12pt]{article}
\usepackage{verbatim}
\usepackage{amsthm}

\usepackage{hyperref}
\hypersetup{
    colorlinks=true,
    linkcolor=blue,
    filecolor=magenta,      
    urlcolor=cyan,
    citecolor=blue,
}

\title{Existence of quantum time crystals}

\author{Franco  Strocchi and  Carlo Heissenberg \\ 
  Scuola Normale Superiore, 56126 Pisa, Italy}
\date{}

\makeglossary

\newtheorem*{Proposition*}{Proposition}

\usepackage{latexsym}
\usepackage[T1]{fontenc}
\usepackage{amsmath}
\usepackage{amssymb}
\usepackage{hyperref}

\def \AO {{\cal A}({\cal O})}
\def \AO' {{\cal A}({\cal O}')}

\def \at {{\alpha_t}}

\def \limx {{\lim_{|\x| \ra \infty}}}

\def \limn  {\lim_{n \ra \infty}}

\def \be {\begin{equation}}
\def \ee {\end{equation}}
\def \ume {{\scriptstyle{\frac{1}{2}}}}
\def \ra {\rightarrow}

\def \eqq {\equiv}

\def \a {{\alpha}}
\def \b {{\beta}}

\def \d {{\delta}}

\def \ph {{\varphi}}

\def \om {{\omega}}

\def \A {{\cal A}}

\def \H {\mbox{${\cal H}$}}

\def \T {{\cal T}}

\def \Z {{\cal Z}}

\def \d^nu {{\partial^\nu}}
\def \d^la {{\partial^\lambda}}
\def \d^o {{\partial^0}}

\def \nbf {{\bf n}}

\def \xbf {{\bf x}}
\def \x {{\bf x}}

\def \Nbf {{\bf N}}

\def \Zbf {{\bf Z}}

\def\doppio#1{{\rm I}\kern-.1667em{\rm #1}}

\def\Q{\text{Q}\kern-.52em
    \text{\vrule height1.5ex width.5pt depth0pt}\kern.45em}

 \def \limN {\lim_{N \ra \infty}}

\def\dZ{{\mathchoice {\hbox{$\Ss\textstyle Z\kern-0.4em Z$}}
{\hbox{$\Ss\textstyle Z\kern-0.4em Z$}} {\hbox{$\Ss\scriptstyle
Z\kern-0.25em Z$}} {\hbox{$\Ss\scriptscriptstyle Z\kern-0.2em
Z$}}}}

\def\dC{{\mathchoice{\hbox{$\rm\textstyle\text{\kern.35em\vrule
   height1.5ex width.5pt depth0pt\kern-.35em C}$}}
{\hbox{$\rm\textstyle\text{\kern.35em\vrule
   height1.5ex width.5pt depth0pt\kern-.35em C}$}}
{\hbox{$\rm\scriptstyle\text{\kern.35em\vrule
   height1.5ex width.3pt depth0pt\kern-.35em C}$}}
{\hbox{$\rm\scriptscriptstyle\text{\kern.35em\vrule
   height1.5ex width.2pt depth0pt\kern-.35em C}$}}}}

\makeatletter

\begin{document}

\maketitle
\begin{abstract}
The existence of quantum time crystals is investigated and shown to be possible in pure phases defined by a state invariant under a group of space translations, as displayed by explicit examples.
\end{abstract}

\section{Introduction}
 The  very innovative  proposal by F. Wilczek \cite{W} to look for quantum time crystals has raised great interest, but their possible existence has been denied \cite{B, WO}. The aim of this note is to partly refine Wilczek proposal, in such a way to overcome the objections of Refs. \cite{B, WO},  and to exhibit explicit examples of quantum time crystals.  

 In our opinion, misleading prejudices affect the issue of possible breaking of time translations. One critical point is the widespread belief that the only criterion of spontaneous symmetry breaking is the non-invariance of a ground state correlation function and  this would a priori preclude the breaking of time translations.    

As a matter of fact, quite generally, a symmetry $\b$ defined as a mapping  of the quantum canonical variables, more generally of (the algebra of) the observables generated by them, fails to define a symmetry of the states of a given realization or phase of the system (i.e. it does not define a Wigner symmetry), if it cannot be described by a unitary operator in the corresponding Hilbert space.

   For an explicit easily checkable criterion of symmetry breaking, in terms of non-invariant expectations,  one has to consider pure states (since invariant  expectations of a mixed state do not exclude symmetry breaking) and therefore  the  characterization of a pure phase as an irreducible (more generally factorial) representation of the algebra of observables enters in a decisive way.  Irreducibility implies that all the operators which commute with the observables must be represented by multiples of the identity. In order to cover the finite temperature case (where irreducibility fails) this condition is only required to apply to the  set of observables which commute with all the observables (briefly the {\em center} of the algebra of the observables);  in this more general case the representation is said to be {\em factorial} and provides the general definition of a pure phase. 

Clearly, in order to relate the breaking of time translations to non-invariant expectations one must consider a state which is not an eigenstate of the (finite volume) Hamiltonian, e.g. a superposition of eigenstates, 
but obviously additional conditions are needed. 

They are easily identified if 
 \,\,i) a group $\T$ of translations is defined on the observables and ii) the following minimal locality condition (asymptotic abelianess)  is satisfied for any pair of observables $A, \,B$ and any translation $T \in \T$
$$ \limn [\,T^n(A), \,B\,] = 0.$$ In the following, we shall always consider pure phases of quantum systems satisfying i) and ii). 

In this case, if a state $\omega$ invariant under $\T$ belongs to the pure phase $\Gamma$, it is the unique $\T$-invariant  state,  equivalently the cluster property holds \cite{FSB}
$$ \limn \langle T^n(A)\,B\rangle = \langle A \rangle\langle B \rangle, $$
where  $\langle C \rangle \eqq \omega(C)$ denotes the expectation of the state $\omega$.

This allows for a simple criterion of spontaneous breaking of a symmetry $\b$ which commutes with the group of translations $\T$ (called an internal symmetry): 

\noindent {\em an internal symmetry $\b$ is spontaneously broken in the pure phase $\Gamma$ if and only if one of the correlation functions of (the translationally invariant state) $\omega$ $\in \Gamma$ is not invariant under} $\b$, 
\be{\langle \b(A) \rangle \neq \langle A \rangle,}\ee equivalently $$\limn \langle T^n (\b(A)) B \rangle  - \langle A \rangle \langle B \rangle \neq 0.$$ 
  In fact, if $\omega$ is invariant under $\b$, by a general result $\b$ may be described by a unitary operator $U_\b$ in the  Hilbert space $\H_\Gamma$ which describes $\Gamma$, i.e. $\b$ defines a Wigner symmetry of $\Gamma$;  conversely, the transformed state state $\omega_\b$, defined by $\omega_\b(A) \eqq \omega(\b(A)) \neq \omega(A)$, is invariant under $\T$ (because so is $\omega$ and $\b$ commutes with $\T$), and by eq.\,(1) and the uniqueness  of the translationally invariant state in $\Gamma$ $\omega_\b$ must  belong to a different phase (and therefore $\b $ cannot be described by a unitary operator). If such a uniqueness property (crucially related to the pure phases) does not hold one cannot  exclude the existence of a unitary operator $U_\b$ which relates the corresponding state vectors $\Psi{\omega_\b} = U_\b \Psi_\omega$, in $\H_\Gamma$.  

It is worthwhile to stress that such a criterion of symmetry breaking, which crucially uses invariance under a group of translations, does not require that $\omega$ is the ground state, even if invariance under a group of translations (not necessarily the full group of translations!) is generically shared  by ground states, and therefore it may be used for the breaking of time translations in a pure phase.      

\section{A mechanism for realizing quantum time crystals. A spin model}  
The next issue is the identification of pure phases defined by a translationally invariant state, which is not invariant under time translations. A possible practical way is to consider a quantum system defined by a Hamiltonian $H$ and a corresponding translationally invariant ground state $\omega$, which defines a pure phase. Then, one switches on an interaction $V$, typically with an external field, formally $H \mapsto H + V$, such that $\omega $ is no longer invariant under the time evolution $\a_t$ defined on the observables  $A$
 by the infinite volume limit of (volume-)cutoffed  Hamiltonian groups 
\be{ \at(A) = \limn e^{i (H + V)_n t} \,A\,e^{-i (H + V)_n t},}\ee
 $n$ indexing the cutoff. The crucial condition is that $\at$ maps observables into observables. 
 In this way, one may  obtain a spontaneous breaking of time translations. 

To explicitly check this,  the following criterion may be useful: 

{\em In a pure phase defined by a state $\om$  invariant under a group of translations  $\T$,  an internal   symmetry $\b$   is spontaneously broken if and only if  there is 
a symmetry breaking order parameter $\om(\b(A)) \neq \om(A)$, for some observable $A$ or, equivalently,  
there is an average or macroscopic  observable
\be{ A_{av} \eqq \limN \frac{1}{N} \sum_{i = 1}^N T^i(A)), \text{ for }T \in \T, }\ee with  non-invariant expectation   $\omega(\b(A_{av})) \neq \omega(A_{av}) = \omega(A)$.} 


Eq.\,(3) immediately implies symmetry breaking, since, by asymptotic abelianess  $A_{av}$ commutes with all the observables so that it is represented by a multiple of the identity in each pure phase and therefore therefore $\b$ cannot be implemented by a unitary operator $U_\b$, which would leave $A_{av}$ invariant. 

It is worthwhile to note that the non-invariance of  expectations  of average/macroscopic observables  on a translationally invariant state is much easier to control, in particular for detecting the breaking of time translations.  


A simple prototype of quantum time crystals is provided by a Heisenberg (lattice) ferromagnet described by a spin one-half Hamiltonian with nearest neighbor coupling $H$. The translationally invariant state $\omega_x$ with all the spins  pointing in the $x$-direction defines an irreducible representation of the observables and a pure phase $\Gamma$, satisfying i) and ii).

Then, one  introduces the interaction $H_1$ with a uniform magnetic field ${\bf h}$ pointing in the $z$-direction. Since $H$ and $H_1$ commute one gets 
\be{ \langle \a_t(\sigma_{x, av}) \rangle  = \cos(|{\bf h}| t) = \langle \sigma^i_{x}(t) \rangle.}\ee
By the above criterion, this implies the breaking of time translations, leaving unbroken the group of time translations with $t = (\ume + n) \pi /|{\bf h}|$, with $\,\,n \in \mathbb Z,$  ({\em quantum time crystal}). For such a conclusion, a crucial  role has been played by the  thermodynamical limit  and by the translational invariance of the state, a point which has not been realized by Watanabe and Oshikawa in their discussion of a simplified version of such a model. Clearly, the impossibility of describing $\a_t$ by a unitary operator in $\H_\Gamma$ prevents the definition of an Hamiltonian there.

The general lesson from this relatively simple model is that if in a given pure phase the average  observables undergo a periodic  motion, one has a breaking of time translations with residual invariance under periodic translations, exactly as it happens for the breaking of space translations in a crystal. 
Thus, the occurence of time crystals does not appear as a very odd, if not impossible, phenomenon as argued in the literature \cite{B,WO}.

\section{An example of time periodicity enforced by topology}
  A similar mechanism may be realized in the model, discussed by Wilczek \cite{W},  of a charged particle with charge $q$ and unit mass confined to a ring  of unit radius that is threaded by (magnetic) flux $2 \pi \alpha/q$.




When $\a = 0$ the model becomes the familiar quantum mechanical model of  a particle on a ring. In the textbook presentations, it is not sufficiently emphasized that the non-trivial topology of the circle has a strong impact on the identification of the algebra of observables $\A$, which is restricted to be the algebra generated by the canonical momentum $p$ and by periodic functions of the angle $\ph$, conveniently by $U((n\ph) \eqq e^{ i n \ph},$ $ n \in \mathbb{ Z}$ \cite{WA}.

 Since the observable $V( 2 \pi) \eqq e^{ i 2 \pi \,p}$ commutes with all the observables, its spectrum,  i.e.  the expectation $\langle V(2 \pi) \rangle_\theta = e^{i \theta}$, $\theta \in [ 0, 2 \pi)$ mod $2 \pi$, labels the  irreducible representations of (the algebra of) the observables hereafter called {\em $\theta$-sectors}, all $\theta$'s being  allowed \cite{Autoagg}.
The mapping  
$$\rho^\theta:    U(n) \,\mapsto \,U(n), \,\,V(\b) \eqq e^{ i \b \,p} \,\mapsto \,V(\b)\,e^{ i  \tilde{\theta}\, \b}, \,\,\,\tilde{\theta} \eqq \theta/2 \pi,$$    defines a symmetry of the observables, which is broken in each $\theta$ sector, since it does not leave the central element $V(2 \pi)$ invariant. The structure is very similar of that of QCD, with $V(2 n \pi)$ playing the role of the large gauge transformations and $\rho^\theta$ the chiral transformations \cite{FS, FS2}.
 
For $\a \neq 0$ the  dynamics $\a^\a(t)$ is given by 
$\a^\a(t) = \rho^{ -2 \pi\,\a}\,\a^0(t)\,\rho^{ 2 \pi\,\a}$, with $\a^0(t)$ the dynamics induced by $H_0 = \ume p^2$. The spectrum of the corresponding Hamiltonian $H^\a$ in each $\theta$ sector coincides with the spectrum of $H^0$ in the sector $\theta - \a$.

As remarked above, to realize a spontaneous symmetry breaking of time translations it is enough to introduce a coupling of $\ph$ with an external field $h$:
$H^{\a,h} =  H^\a + h \ph.$
Such a Hamiltonian does not belong to the algebra of observables, but nevertheless the corresponding time evolution $\a^{\a,h}(t)$ maps the algebra of observables $\A$ into itself, as required,  for physical consistency.
In fact, by using  Zassenhaus' formula, one easily gets 
 $$\a^{\a, h}(t)(U(n)) = e^{- i \ume  t^2 h (p - \a)}\,\a^\a(t)(U(n))\, e^{ i \ume  t^2 h (p - \a)} \in \A,$$
 \be{\a^{\a, h}(t)(V(\b)) = V(\b)\,e^{i \b\, h\,t}, \,\,\,
\a^{\a, h}(t)(V(2 \pi)) = V(2 \pi)\,e^{i 2 \pi h\,t}.}\ee 
Hence, $\a^{\a, h}(t)$ is broken in each $\theta$ sector, leaving  as residual invariance group the time translations with $t = n/h, \,n \in \mathbb Z$.

The non-trivial topology of the circle plays a crucial role for the symmetry breaking in such a finite dimensional model; in the (reducible) representation with a decompactified $\ph$, given by $L^2(\mathbb R, d \ph)$, the dynamics  $\a^{\a, h}(t)$ is implemented by unitary operators, which, however, do not leave each $\theta$ sector invariant.     

\section{Breaking of time translations in many-particle Wilczek model}
For a many particle extension of the model discussed above, consider $N$ copies of the Wilczek model, e.g. by considering $N$ coaxial rings with the same unit radius, labeled by the index $i$,  with the following Hamiltonian:
$ H^{\a, V, N} = \sum_{i = 1}^N( H^\a_i + V(\ph_i)),$
with $V(\ph_i)$ a periodic function, to guarantee that  the corresponding time evolution $\a^{\a, V, N}(t)$ maps observables into observables \cite{WF}. 

In order to find phases with broken time translations in the $N \ra \infty$  limit, consider the product state  $\Psi_0^N \eqq \prod_{i = 1}^N \,\Psi_0^i$, with $\Psi_0^i = (2 \pi)^{-\ume} \,e^{i \a \ph_i}$ the ground state of $H^{\a, i}$ (one might as well take the product of the same $\theta$ state for each index $i$). Clearly, in the thermodynamical limit such a state defines a pure phase and it is invariant under the group of ``translations'' of the index $i$, so that the criterion of Section 2 applies. 

To this purpose, one may consider  
 the following observable
$$ P_N = \frac{1}{N} \sum_{i = 1}^N f(\ph_i)\,p_i\,f(\ph_i) \eqq \frac{1}{N}\sum_{i = 1}^N p_i^f,$$    
with $f$ a regular function of compact support in $[0, 2 \pi]$, with periodic extension $f(\ph_i + 2 \pi) = f(\ph_i)$; the observable $p_i^f$ describes a momentum localized in the support of $f$ and its expectation on the corresponding  $\Psi_0^i $ is $\langle p_i^f \rangle_i = \a\,(2 \pi)^{-1} \int _0^{2 \pi} d \ph_i\,|f(\ph_i)|^2 \neq 0$, and is independent of the index $i$. 

 Furthermore, denoting by $\a^{\a, V, N}(t)$ the time evolution defined by $H^{\a, V, N}$ and by $\langle \ldots \rangle_{\a, N}$ the expectation on the above product state, one has 

$$ \frac{d}{d t} \langle \a^{\a, V, N}(t)(p_i^f) \rangle_{\a, N}|_{t = 0} =  \int _0^{2 \pi} \frac{d \ph_i}{2 \pi} \,|f(\ph_i)|^2 \,\frac{d V(\ph_i)}{d \ph_i},$$
having used that $\Psi_0^i$ is the ground state  of the kinetic Hamiltonian $H_i^{\a}$. For a given  potential one can always find $f$ such that  the right hand side does not vanish (thanks to the fact that $\Psi_0^i$ is not an eigenstate of $H_i^{\a, V}$). 

As in the example of spin systems, in the  infinite particle limit the  average  observable
$$P_{av} = \limN \frac{1}{N} \sum_{i = 1}^N p_i^f$$  commutes with  the observables, it is not zero in the (factorial) representation defined by the above product state in the limit $N \ra \infty$, $\langle P_{av} \rangle_{\a, \infty} = \langle p_i^f \rangle_{\a, \infty}$,   and 
\be{ d/dt \langle   \a^{\a, V, \infty}(t)(P_{av}) \rangle_{\a, \infty}|_{t = 0}  \neq 0.}\ee
Thus, such an average  observable has a non-trivial time evolution and therefore time translations are broken 
 in such a  representation.

Another possibility is to consider the average/macroscopic observable 
$$ F(\ph)_{av} \eqq \limN \frac{1}{N} \sum_{i=1}^N F(\ph_i),$$
with $F$ a regular periodic function of compact support.

Proceeding as before one gets 
$$d/dt \langle \a^{\a, V, N}(t)(F(\ph_i)) \rangle_{\a, N}|_{t = 0} = 0, $$
\be{(d/dt)^2 \langle \a^{\a, V, N}(t)(F(\ph_i)) \rangle_{\a, N}|_{t = 0} = -  \langle V'(\ph_i)\,F'(\ph_i) \rangle_{\a, N },
}\ee where the prime denotes the derivative with respect to $\ph_i$. The right hand side is independent of $i$ and different from zero for all periodic functions $F$, with derivative which is  not orthogonal to $V'$;  therefore the same result holds for $F(\ph)_{av}$ in the $N \ra \infty$ limit. By the above criterion, this implies the breaking of time translations.  

The conclusions do not change if the interaction $V$  also contains a particle-particle interaction term $V_{p-p} = \sum_{i<j} v(\ph_i - \ph_j)$, with $v$ a periodic function, since this term does not contribute to the right hand sides of eqs.\,(6, 7).

F. S. is grateful to Giovanni Morchio for very useful discussions.


\begin{thebibliography}{9}
\bibitem
{Autoagg} The corresponding Hilbert space representation is given either by $\pi_\theta(p) = - i\, \partial/\partial \ph + \theta/ 2 \pi$ acting on periodic functions or, equivalently, $\pi_\theta(p) = - i\, \partial/\partial \ph$ acting on quasi periodic functions such that $\psi(2 \pi) = e^{i \theta}\, \psi(0)$.
\bibitem
{WF} The dynamics defined by the Hamiltonian of eq.\,(13), in Wilczek's paper \cite{W}, does not seem to satisfy this condition.
\end{thebibliography}
\end{document}